\documentclass{aa}  

\usepackage{graphicx}
\usepackage{hyperref}
\usepackage{xcolor}
\usepackage{mathabx}
\usepackage{footmisc}
\usepackage{txfonts}
\begin{document}

   \title{Volatile atmospheres of lava worlds\thanks{The code and data used to produce the results presented in this study are available in the branch {\it lava\_worlds} of the git repository of the code MOAI (\href{https://bitbucket.org/MaximeMaurice/moai/src/Maurice_etal_2024/}{https://bitbucket.org/MaximeMaurice/moai/src/Maurice\_etal\_2024/}).}}


   \author{M. Maurice
          \inst{1,2}
          \and
          R. Dasgupta\inst{2}
          \and
          P. Hassanzadeh\inst{2,3}
          }

   \institute{Laboratoire de Météorologie Dynamique, IPSL, CNRS,\\
              4 Place Jussieu, 75252 Paris, France\\
              \email{maximemaurice@protonmail.com}
              \and
              Rice University, Earth, Environmental and Planetary Sciences Department,\\
              6100 Main Street, MS 126, Houston, TX 77005, USA \\
              \and
              Rice University, Mechanical Engineering Department,\\
              6100 Main Street, MS 126, Houston, TX 77005, USA \\
             }

   \date{Received XXX; accepted YYY}

 
  \abstract
   {A magma ocean (MO) is thought to be a ubiquitous stage in the early evolution of rocky planets and exoplanets. During the lifetime of the MO, exchanges between the interior and exterior envelopes of the planet are very efficient. In particular, volatile elements that initially are contained in the solid part of the planet can be released and form a secondary outgassed atmosphere.}
   {We determine trends in the H-C-N-O-S composition and thickness of these secondary atmospheres for varying planetary sizes and MO extents, and the oxygen fugacity of MOs, which provides the main control for the atmospheric chemistry.}
   {We used a model with coupled chemical gas-gas and silicate melt-gas equilibria and mass conservation to predict the composition of an atmosphere at equilibrium with the MO depending on the planet size and the extent and redox state of the MO. We used a self-consistent mass-radius model for the rocky core to inform the structure of the planet, which we combined with an atmosphere model to predict the transit radius of lava worlds.}
   {The resulting MOs have potential temperatures ranging from 1415 to 4229 K, and their outgassed atmospheres have total pressures from 3.3 to 768 bar. We find that MOs (especially the shallow ones) on small planets are generally more reduced, and are thus dominated by H$_2$-rich atmospheres (whose outgassing is strengthened at low planetary mass), while larger planets and deeper MOs vary from CO to CO$_2$-N$_2$-SO$_2$ atmospheres, with increasing $f_{{\rm O}_2}$. In the former case, the low molecular mass of the atmosphere combined with the low gravity of the planets yields a large vertical extension of the atmosphere, while in the latter cases, secondary outgassed atmospheres on super-Earths are likely significantly shrunk. Both N and C are largely outgassed regardless of the conditions, while the S and H outgassing is strongly dependent on the $f_{{\rm O}_2}$, as well as on the planetary mass and MO extent for the latter. We further use these results to assess how much a secondary outgassed atmosphere may alter the mass-radius relations of rocky exoplanets.}
   {}


   \maketitle
%

\section{Introduction}
\label{sec:intro}
Large-scale melting of the silicate mantles of terrestrial planets is common during the late phases of their growth \citep{Stevenson1981,Wood2006,ElkinsTanton2012,Chao2021}. These episodes of magma ocean (MO) give rise to extensive chemical equilibration between the silicate reservoir of the planet and its exterior, allowing the outgassing of volatile elements that are originally present in its building blocks \citep{AbeMatsui1988,ElkinsTanton2008}, or the ingassing of nebular material \citep{OlsonSharp2019}. The variability in solubility exhibited by the different volatile species leads to different outgassing patterns for different elements \citep{ElkinsTanton2008,Lebrun2013,Bower2019,Nikolaou2019,Lichtenberg2021}, further altered by redox-sensitive speciation \citep{Katyal2020,Bower2022,Gaillard2022}. For modest atmospheric masses ($\ll1$\% of the planetary mass) that are typical of rocky planets, the bulk redox state of the MO-atmosphere system  is set by core-mantle equilibration during the planetary differentiation and is quantified by the oxygen fugacity ($f_{{\rm O}_2}$).
In the Solar System, the terrestrial MO likely outgassed a neutral to oxidized atmosphere \citep{Hirschmann2012,Sossi2020,Deng2020,Kuwahara2023} due to the significant $f_{{\rm O}_2}$ gradient between the MO bottom, where $f_{{\rm O}_2}$ is buffered by equilibration with Fe alloy, and its surface, which in turn buffers the redox state of the atmosphere. Conversely, smaller bodies have been theorized to outgas at more reducing conditions \citep{Deng2020,Armstrong2019} because they lack such a pressure-induced negative gradient  and because the $f_{{\rm O}_2}$ pressure-gradient is positive at shallow depths.

Beyond the Solar System, rocky exoplanets likely orbit most stars in the Milky way, and the catalog of observations is growing. With the current limitations of rocky exoplanet observations, the radius of rocky exoplanets ranges to about 1.6 Earth radii (the lower bound of the so-called radius valley; \citep{Fulton2017}), corresponding to approximately 6 Earth masses. The outgassed atmospheres during MO episodes on exoplanets probably exhibit a diversity that reflects their sizes, compositions and chemistries (e.g., redox state). The fate of highly volatile elements that are essential for life (H-C-N-O-S) is strongly influenced by the MO outgassing because once they are delivered to the atmosphere, these elements are prone to escaping to space, especially during the pre-main sequence of the host star, when it emits a strong X-UV flux that can drive hydrodynamic escape \citep{Lammer2018}. Understanding MO outgassing is thus key to any theory of planetary habitability. Furthermore, atmospheric characterization by ongoing and upcoming missions will soon give access to the chemical composition of atmospheres of terrestrial exoplanets. The origin of these atmospheres can be diverse (nebular accretion, volcanic outgassing, delivery of volatile-rich cometary materials, etc.; \citep{KrisanssenTotton2020,KiteBarnett2020}), and MO outgassing provides a potentially important source mechanism that is particularly relevant for young systems. Therefore, observations need to be accompanied by the most advanced models in order to understand the planetary context of the studied exoplanets. Several studies have investigated the composition of MO-outgassed volatile atmospheres relevant to early Earth or Mars \citep{ElkinsTanton2008,Lebrun2013,Salvador2017,Nikolaou2019,Lichtenberg2021}, as well as the effect of the redox state of the planet \citep{Katyal2020,Bower2022,Gaillard2022}, but a systematic study of the planetary parameters relevant for the diversity of exoplanets, in particular, linking these parameters with the oxidation state of the atmosphere, is still needed.

While some outgassing trends can be derived from simple physical principles, the chemical speciation of the atmosphere can significantly complicate the picture. Hence, in this work, we use an MO outgassing chemical model to calculate the composition of outgassed H-C-N-O-S atmospheres on MO exoplanets as a function of their mass and of the extent of their MO (more specifically, the molten silicate mass fraction, i.e., the ratio of the mass of the MO to the total mass of silicate, which we denote as $\Phi$), and of their redox state. We characterize the extent of the outgassing of each element and the gaseous speciation in the atmosphere. These results are used to infer the possible gas-depletion patterns of exoplanets (i.e., the relative depletion in one volatile element compared to another in the bulk planet) and to predict observables in terms of alterations in the mass-radius relations for MO exoplanets.

\section{Model}
\label{sec:model}
We first determine the mass-radius relations of a set of planets with various masses and $\Phi$. This allows us to calculate the mass of the MO, its bottom pressure, and its  profile (all necessary quantities for the chemical model). We then compute the surface $f_{{\rm O}_2}$ based on the core-mantle equilibrium $f_{{\rm O}_2}$ ($f_{{\rm O}_2,{\rm eq}}$) at the bottom of the MO, which we vary as another free parameter. Finally, we use a chemical equilibrium model coupled to volatile mass conservation to calculate the atmospheric pressure and composition of these planets.

\subsection{Mass-radius relation of refractory reservoirs}
\label{subsec:planet_core}
\begin{figure*}
    \centering
    \includegraphics[width=\textwidth]{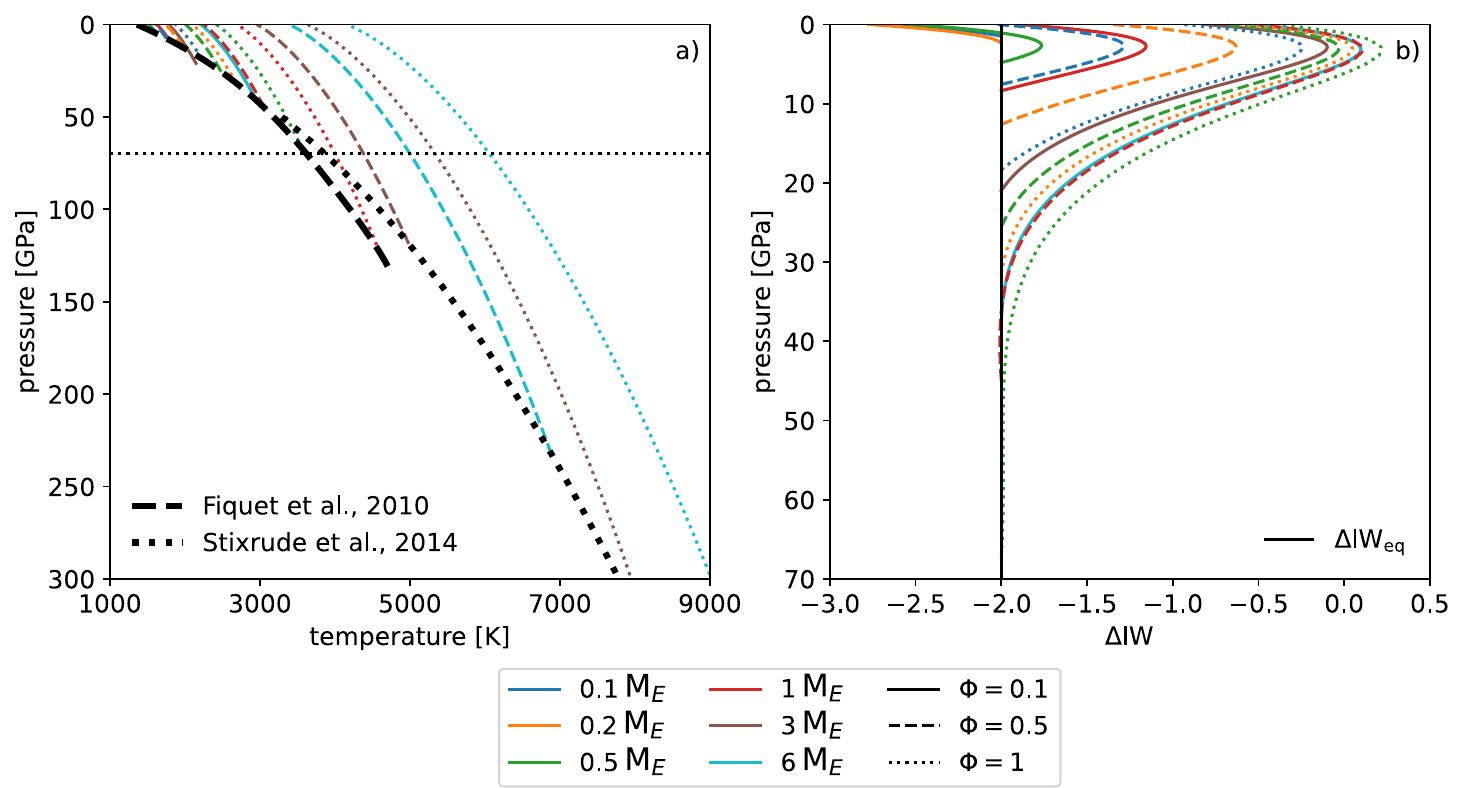}
    \caption{Vertical profiles of temperature (a) and $f_{{\rm O}_2}$ (noted as the log-unit difference from the IW buffer, b) through the MO for all $M_p$ and $\Phi$, for $\Delta{\rm IW}_{\rm eq}=-2$. The thick dashed and dotted black lines in panel (a) correspond to the solidus from \citet{Fiquet2010} (used for Earth and sub-Earth-mass planets), and the solidus from \citet{Stixrude2014} (used for super-Earth-mass planets), respectively. The solidus defines the temperature reached at the bottom of the MO. The thin dotted black line corresponds to the maximum pressure in panel (b). The solid vertical line in panel b corresponds to $\Delta{\rm IW}_{\rm eq}=-2$, which is the value reached at the bottom of the MO. The profile of $\Delta$IW for cases with $\Delta{\rm IW}_{\rm eq}=0$ ($\Delta{\rm IW}_{\rm eq}=-5$) are offset by 2 ($-3$) log-units on the $x$-axis of panel (b).}
    \label{fig:profiles}
\end{figure*}

There exists a wealth of literature on rocky exoplanets mass-radius relations \citep{Valencia2006,Dorn2015,Zeng2016,Zeng2019,Otegi2020,Agol2021,Unterborn2016,Unterborn2023}. However, to the authors' knowledge, only \citet{DornLichtenberg2022} took into consideration the density difference between solid and (hot) liquid silicate in their mass-radius relations.
We considered a set of planets of masses ($M_p$) between 0.1 and 6 Earth masses with a molten silicate mass fraction ($\Phi$) between 0.1 and 1. For each set of parameter values, we used burnman \citep{burnman} to generate a three-layer planet composed of pure fcc iron (using the corresponding equation of state of \citet{SaxenaErikson2015}), a bridgmanite solid lower mantle (using the corresponding equation of state from \citet{deKokerStixrude2013}), and a liquid MgSiO$_3$ MO (using the corresponding equation of state from \citet{deKokerStixrude2013}). While the smallest planets (approximate mass of Mars) are unlikely to have stable bridgmanite, and the melting curves we used to compute the depth of the MO are matched for a natural peridotite composition, we kept the same equation of state in order to isolate the effect of the outgassed atmosphere on the mass-radius relations discussed below. For all these planets, we adjusted the core-mantle boundary, MO bottom (when different), the planetary radii ($R_{\rm CMB}$, $R_{\rm MO}$, and $R_p$, respectively), and the potential temperature of the MO ($T_{\rm MO}$), to recover 1) the desired total mass and molten silicate mass fraction, 2) an Earth-like iron core mass fraction (0.32), and 3) to match the solidus with the MO adiabat at the MO bottom. We used the solidus from \citet{Fiquet2010} for Earth- and sub-Earth-sized planets and that from \citet{Stixrude2014} for super-Earths. Criteria 1--3 are fulfilled with a precision of 1\%. All reservoirs were assumed to be adiabatic, and the thermal boundary layers between them were ignored.
The resulting geometries are indicated in Table \ref{tab:planetary_cores}, and the temperature profiles are represented in Figure \ref{fig:profiles}a.

\subsection{MO redox state}
\label{subsec:MO_redox_state}
The process resulting in an MO (e.g., giant impact) is generally associated with an event of alloy-silicate equilibration (the alloy originating from the core of the impactor in the case of a giant impact). This equilibrium buffers the redox state of the MO at a certain value of $f_{{\rm O}_2,{\rm eq}}$ depending on the compositions of the alloy and silicate melts \citep{Rubie2015}. This value, expressed as $\Delta$IW$_{\rm eq}$, that is, the difference in log-units from the iron-w\"ustite (IW) buffer at the same $P$-$T$ conditions, was treated as a free parameter and was varied between 0 and $-5$, roughly encompassing values estimated for the Solar System \citep{Rubie2015}. $f_{{\rm O}_2,{\rm eq}}$ sets the ${\rm Fe}^{3+}/\Sigma{\rm Fe}$ ratio in the MO, where the equilibrium between ferrous and ferric iron oxides is the redox buffer after scavenging of Fe alloy in the core \citep{Hirschmann2012,Armstrong2019,Deng2020,Sossi2020,Hirschmann2022}. We calculated this ratio considering that our chosen $f_{{\rm O}_2,{\rm eq}}$ was reached at the bottom of the MO where we assumed that the alloy-silicate equilibrium took place. From the obtained ${\rm Fe}^{3+}/\Sigma{\rm Fe}$ (assumed to be homogeneous in the MO) and the MO $P$-$T$ profile (Figure \ref{fig:profiles}a), we computed the $f_{{\rm O}_2}$ profile through the MO using the parameterization from \citet{Hirschmann2022} and \citet{Deng2020} (represented in Figure \ref{fig:profiles}b). The $f_{{\rm O}_2}$ reached at the surface is the value at which the atmosphere is buffered. 

It has been shown that the slope of the $f_{{\rm O}_2}$ profile in the MO changes sign between 5 and 10 GPa, that is, the MO becomes oxidized with increasing depth at shallow levels ($P<\sim6$ GPa, \citep{ONeill2006,Zhang2017}) and then starts to be reduced with increasing depth at deeper levels ($P>\sim6$ GPa) \citep{Hirschmann2012,Deng2020,Sossi2020,Kuwahara2023}. A consequence of this is that as the planet becomes larger and/or the MO becomes deeper, beyond 5-10 GPa, the surface of the MO becomes increasingly oxidized (Fig. \ref{fig:profiles}). None of the existing parameterizations is calibrated at pressures higher than 55 GPa \citep{Deng2020}. However, the pressure gradient of the $f_{{\rm O}_2}$ may cease to exist at large depths (Figure \ref{fig:profiles}b), and we therefore considered all MOs with a bottom deeper than 55 GPa to have a surface $f_{{\rm O}_2}$ that corresponds to the highest value reached for MOs shallower than 55 GPa, which is $\sim$ 1.5 log-units above the bottom value $\Delta$IW$_{\rm eq}$.

We kept a constant core-mass ratio while varying the redox state at which the core and the MO equilibrate. However, these two quantities are not independent: A larger core generally implies a more reduced mantle \citep{Rubie2015}. Nevertheless, a straightforward relation only holds for a single-stage core-mantle equilibration, which is a highly unrealistic scenario. In addition, other processes (e.g., an evolution in the composition of the accreting material) can induce a deviation in the redox state of the MO from that imposed by equilibration with the bulk core. Therefore, bearing this caveat in mind, we chose to vary the redox state of the MO as a free parameter while keeping a constant, Earth-like core mass ratio, similar to that which is most frequently represented in rocky planets of the Solar System.

\subsection{MO atmosphere equilibrium}
\label{subsec:equilibrium}
The chemical model computes the partial pressures of the various gaseous species in the atmosphere and the elemental solubilities based on the mass of the MO, the radius, and the gravity at the surface of the planet, a set of $T-$dependent equilibrium constants (evaluated at the potential temperature of the MO), and the bulk masses of H, C, N, and S present in the MO-atmosphere system. It accounts for mass conservation of all four volatile elements, present under 11 different gaseous species in the atmosphere (H$_2$, H$_2$O, CO, CO$_2$, CH$_4$, N$_2$, NH$_3$, HCN, SH$_2$, S$_2$, and SO$_2$), dissolution equilibria between gas and silicate melt, and the seven following gas-gas redox equilibria:
\begin{eqnarray}
    &{\rm H}_{2}+1/2{\rm O}_{2}&={\rm H}_2{\rm O}_{}\label{eq:H_equi} \\
    &{\rm CO}+1/2{\rm O}_{2}&={\rm CO}_{2}\label{eq:C_equi} \\
    &{\rm CH}_{4}+{\rm O}_{2}&={\rm CO}+2{\rm H}_2{\rm O}\label{eq:CH_equi} \\
    &{\rm N}_{2}+3{\rm H}_{2}&=2{\rm NH}_{3}\label{eq:N1_equi} \\
    &{\rm H}_2{\rm O}+{\rm HCN}&={\rm CO}+{\rm NH}_{3}\label{eq:N2_equi} \\
    &{\rm S}_{2}+2{\rm O}_{2}&=2{\rm SO}_{2}\label{eq:S1_equi} \\
    &2{\rm H}_2{\rm S}&={\rm S}_{2}+2{\rm H}_{2}\label{eq:S2_equi}.
\end{eqnarray}
The system of equations that is solved consists of the seven chemical equilibria above, and the mass conservation of H, C, N and S. We note that O mass conservation is omitted and replaced with the buffering of $f_{{\rm O}_2}$ by the MO, as described in the previous section. The general form of the chemical equilibria equations is
\begin{equation}
    \prod_{s} f_s^{c_s}=K(T),
\end{equation}
where $f_s$ is the fugacity of the gaseous species $s$ at the surface, $c_s$ is the signed stoichiometric coefficient (positive for products, negative for reactants) of the gaseous species $s$ in the equilibrium under consideration, $K$ is the equilibrium constant of the equilibrium under consideration, and $T$ is the potential surface of the MO (assumed to coincide with the surface temperature). The fugacities are calculated as the partial pressures divided by a reference pressure of one bar. The equilibrium constants of Eqs \ref{eq:H_equi}--\ref{eq:S2_equi} are calculated from each species' $T-$dependent formation Gibbs-free energy tabulated from the JANAF tables \citep{JANAF}.

The general form of the chemical equilibrium equations (derived in Appendix \ref{apx:mass_cons}) is
\begin{equation}
    M_e=\sum_s\left[\left(M_{\rm MO}S_{s}(p_{s},P,f_{{\rm O}_2})+\frac{4\pi R_{p}^2}{g}p_{s}\frac{\mu_{s}}{\mu_{\rm atm}}\right)\frac{\mu_{e}\lambda_s^e}{\mu_{s}}\right],
    \label{eq:mass_cons}
\end{equation}
where $M_e$ is the bulk mass of element $e$, $M_{\rm MO}$ is the mass of the MO; $S_s$, $p_s$, and $\mu_s$ are the solubility law, partial pressure at the surface, and molecular weight of species $s$, respectively; $P$ and $f_{{\rm O}_2}$ are the total pressure and the oxygen fugacity at the surface; $R_p$ and $g$ are the planetary radius and surface gravity, respectively; $\mu_{\rm atm}$ is the average molecular weight in the atmosphere; $\mu_e$ is the atomic weight of element $e$; and $\lambda_s^e$ is the number of atoms of $e$ in one molecule of $s$. 
In equation \ref{eq:mass_cons}, the first term in the innermost brackets corresponds to the mass of species $s$ dissolved in the MO, and the second term corresponds to the mass present in the atmosphere. The term in the outermost bracket converts the mass of species $s$ into the mass of element $e$, and the total is summed over all $e$-bearing species. We benchmarked our outgassing model against other similar models from the literature \citep{Bower2022,Gaillard2022} for the cases presented in \citet{Bower2022} (see Appendix \ref{apx:benchmark}).

In order to maintain a manageable parameter space, we restricted our study to a fiducial case with bulk contents of 100 ppm of the MO mass for each element. The volatile delivery during planet formation is not well understood, and the observed fractionation between these elements could be inherited from the building blocks \cite[e.g.][]{Piani2020,Grewal2021} or might postdate the MO phase \citep{Albarede2009}. In addition, H ingassing from the protoplanetary disks (for MOs on planets that are still embedded in the primordial nebula) could also alter the H budget, with an increasing effect of planetary mass \citep{OlsonSharp2019,Young2023}. \citet{Stoekl2016} calculated that in a solar nebula-like disk, a Mars-sized planet could attract an H$_2$ envelope of $\sim$100 ppm of its bulk mass by the time of disk dispersal (1 to 10 Myr), which corresponds to our fiducial value for H. Larger planets are likely to accrete more nebular gas (\citet{Stoekl2016} found that super-Earths are likely to trigger a gravitational runaway phase, but their model provided an upper bound on accretion because it neglected erosion processes). We therefore extended our parameter space to a higher H budget (1000 ppm) in order to investigate this possibility. Finally, as rocky bodies in the Solar System exhibit a consistent N depletion compared to C and H \citep{Marty2012,Bergin2015,DasguptaGrewal2019}, we also tested cases in which the N budget was decreased to 10 ppm.

\subsection{Volatile solubilities}
\label{subsec:volatiles_solubilities}
%
Unless stated otherwise, the solubilities were set by gas-melt equilibria. The corresponding species-specific solubility laws, their references, and their calibration domains are reported in Table \ref{tab:solubilities}: We used Henry law coefficients from \citet{Hirschmann2012b} for H$_2$ and those of \citet{Hirschmann2016} for CO, CO$_2$, and for CH$_4$ (C solubility under ``oxidized'', ``reduced'', and ``very reduced conditions'' in their Table 2, respectively), and the recent parameterization for water solubility in peridotitic melts by \citet{Sossi2023}. We neglected dissolution of HCN, NH$_3$, SH$_2$, and SO$_2$ and used the $f_{{\rm O}_2}$-dependent solubility laws from \citet{Dasgupta2022} for N$_2$ and from \citet{Gaillard2022} for S$_2$. Hence, the elemental solubilities of all elements are sensitive to $f_{{\rm O}_2}$: H and C because of the different solubilities of their redox end-member carriers, and N and S through a direct $f_{{\rm O}_2}$-sensitive solubility. It is important to note that these various parameterizations have been calibrated by laboratory experiments in a parameter space that sometimes does not extend to actual MO-like conditions. In particular, the upper bound in temperature at which these experiments are performed corresponds to the lower range of the temperatures envisaged here (see Table \ref{tab:planetary_cores}). Hence, high-temperature experiments, although challenging, are needed to calibrate models such as the one used here, and to constrain the influence of temperature on volatile solubility. Melt composition is also a limitation: Most experiments (except for \citet{Sossi2023}) were performed on basaltic composition, which is relevant for volcanic outgassing, but is not accurate for bulk rocky reservoirs.

It is possible that a given volatile element, rather than being either dissolved in the MO or outgassed to the atmosphere, is also incorporated in a condensed phase. Based on the solubilities, this possibility is highest for graphite/diamond precipitation. To account for the possibility of graphite precipitation, we calculated the CO$_2$ and CO concentrations at graphite saturation based on the $P-T-f_{{\rm O}_2}$ profiles in the MO (Figure \ref{fig:profiles}) using the parameterizations from \citet{EguchiDasgupta2018} and \citet{Yoshioka2019}, respectively. Assuming a well-mixed MO, the concentration of each species is constant throughout the MO, and graphite will precipitate if the C concentration exceeds the minimum value of the respective saturation profiles. If this occured, we decreased the bulk C mass by 1\% and computed the new chemical equilibrium. We iterated until the CO$_2$ and CO concentrations in the MO were both lower than or equal to their minimum saturation values in the MO. The CO-CO$_2$ content of the MO was thus buffered at the graphite saturation at the bottom of the MO due to the decreasing solubility of these species with pressure (at equilibrium with graphite). As these parameterizations have only been calibrated up to a few GPa and $\sim1500$ K and yield virtually zero solubility when extrapolated at higher pressures, we only tested the shallowest MO cases ($\Phi=0.1$, $M_p\leq0.5$ Earth mass).

\section{Results}
\label{sec:results}

\subsection{Physics of outgassing}
\label{subsec:extent_outgassing}
The influence of the planetary mass and molten silicate mass fraction ($\Phi$) on outgassing is best understood for a simplified case, where a single volatile element is present as a single gaseous species, obeying a Henrian solubility law (i.e., the molar concentration of volatile in the MO is proportional to the partial pressure of the species, which itself is equal to the total atmospheric pressure $P$). 
In a single-species atmosphere, considering a Henrian solubility law, the ingassed and outgassed masses are
\begin{eqnarray}
    & M_{\rm in} &= M_{\rm MO}\alpha P, \label{eq:M_in}\\
    & M_{\rm out} &= \frac{4\pi R_p^2}{g}P=\frac{4\pi R_p^4}{GM_p}P, \label{eq:M_out} 
\end{eqnarray}
where $G$ is the universal gravity constant, and $\alpha$ is the Henrian solubility constant. For a fixed $\Phi$, we have $M_{\rm MO}\propto M_p$, and the out-to-ingassed mass ratio is therefore proportional to $R_p^4/M_p^2$, which is a decreasing function of $R_p$ for any realistic mass-radius relation (e.g., for a homogeneous sphere where $M_p\propto R_p^3$, or also for more realistic cases: $M_p\propto R_p^{1/0.27}$ \citep{Valencia2006} or $M_p\propto R_p^{1/0.29}$ \citep{Otegi2020}).

In order to ensure mass conservation, $P$ increases as $M_p$ increases (at a fixed $\Phi$ and bulk volatile concentration). Summing \ref{eq:M_in} and \ref{eq:M_out}, we can write
\begin{equation}
    P=\frac{M_{\rm bulk}}{M_{\rm MO}\alpha+\frac{4\pi R_p^2}{g}}, \label{eq:P}
\end{equation} 
where $M_{\rm bulk}=M_{\rm in}+M_{\rm out}$ is the bulk mass of volatiles in the MO+atmosphere system. Taking the limit of \ref{eq:P} for low $M_p$ yields $P\sim M_{\rm bulk}g/(4\pi R_p^2)$, which is the expression for the pressure of an atmosphere of mass $M_{\rm bulk}$, thus corresponding to a fully outgassed case. Conversely, taking the limit of \ref{eq:P} for high $M_p$ yields: $P\sim M_{\rm bulk}/(\alpha M_{\rm MO})$, that is, $\alpha P\sim M_{\rm bulk}/M_{\rm MO}$, which is the expression of the solubility ($\alpha P$) when volatiles are completely ingassed (replacing $M_{\rm in}$ with $M_{\rm bulk}$ in \ref{eq:M_in}).

If there are multiple species in the atmosphere, Eq. \ref{eq:P} becomes
\begin{equation}
    p=\frac{M_{\rm bulk}}{M_{\rm MO}\alpha+\frac{4\pi R_p^2\mu/\mu_{\rm atm}}{g}}, \label{eq:P2}
\end{equation} 
where $p$ is the partial pressure of the species under consideration, $\mu$ is its molecular mass, and $\mu_{\rm atm}$ is the mean molecular mass of the atmosphere. If $\mu_{\rm atm}$ decreases (e.g., because a light species is added), then $p$ decreases as well. This it does not mean that the species is being ingassed, however: As $p$ decreases, the ingassed mass ($M_{\rm MO}\alpha p$) decreases as well. On the contrary, the positive effect of the decrease of $\mu_{\rm atm}$ on the outgassed mass dominates the negative effect of the decrease of $p$: The species is being outgassed although its partial pressure decreases. The opposite occurs when $\mu_{\rm atm}$, and addition of heavy species leads to net ingassing.

Finally, keeping planetary parameters (mass, gravity, and radius) constant, the outgassed reservoir only varies with $P$, while the ingassed reservoir varies as $\Phi\times P$. The ingassed-to-outgassed mass fraction thus varies as $\Phi$. Again, the atmospheric pressure has to increase so that the in- and outgassed mass fractions add up to one.

\begin{figure*}
    \centering
    \includegraphics[width=\textwidth]{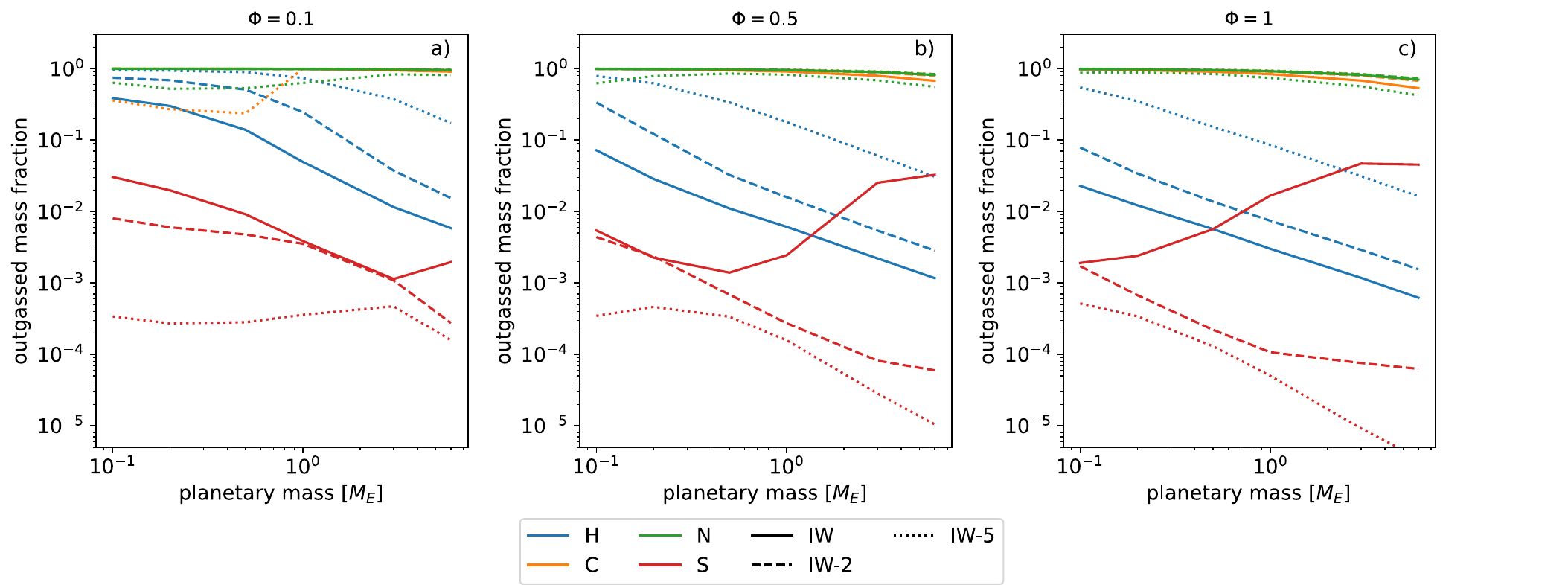}
    \caption{Outgassed mass fraction of all volatiles as a function of planetary mass for a shallow MO (a), a half-mantle MO (b), and a whole-mantle MO (c). The curve colors and style correspond to element and $\Delta$IW$_{\rm eq}$, respectively.}
    \label{fig:outgassed_mass_fraction}
\end{figure*}

We verified these trends by monitoring the outgassed mass fraction of each element upon varying the planetary mass (Figure \ref{fig:outgassed_mass_fraction}). C-bearing species have very low solubilities compared to most other species (except for the most reduced cases, where graphite precipitates). They remain mostly in the atmosphere, although their outgassed fraction decreases slightly at high planetary masses and molten silicate mass fraction. In the most reduced cases (for 0.1 and 0.5 $M_E$ with $\Phi=0.1$; Figure \ref{fig:outgassed_mass_fraction}a) graphite precipitates, significantly decreasing the outgassed mass fraction. While we only accounted for graphite saturation in the smallest planets with shallow MOs, this probably also applies to other reduced cases, and our result could overestimate the amount of outgassed carbon in other reduced cases. Unfortunately, the conditions for graphite precipitation at higher $P-T$ remain unconstrained.
N solubility follows a similar trend (although slightly less pronounced). However, at a low planetary mass, it features a decrease that is not explained by these physical considerations but by the $f_{{\rm O}_2}-$dependence on the N solubility (see Section \ref{subsec:composition}).
Because the elemental solubility of H is offset by the very high solubility of water, its outgassed mass fraction is always much smaller. However, it also follows a clear trend of a decreasing outgassed mass fraction with increasing planetary mass. The water solubility does not follow a Henrian behavior, but still induces the same trend, as observed by \citet{Sossi2023}. Finally, while this trend is also followed by S for shallow and/or reduced MOs (Figure \ref{fig:outgassed_mass_fraction}a), the decreasing solubility of S with increasing $f_{{\rm O}_2}$ (similar to N) reverses this trend as $\Delta$IW$_{\rm eq}$ and $\Phi$ increase. In general, decreasing outgassed fractions with increasing planetary mass suggest that the volatile budget of super-Earths is much less prone to loss due to MO outgassing and subsequent atmospheric escape than smaller planets, in particular, for H \citep{Sossi2023}. Conversely, small planets such as Mars outgas a large fraction of their H, which is then prone to escaping due to the low binding energy of these planets (see Section \ref{subsec:fate_volatiles}). 

\subsection{Chemistry of outgassing}
\label{subsec:composition}
\begin{figure}
    \centering
    \includegraphics[height=0.6\textheight]{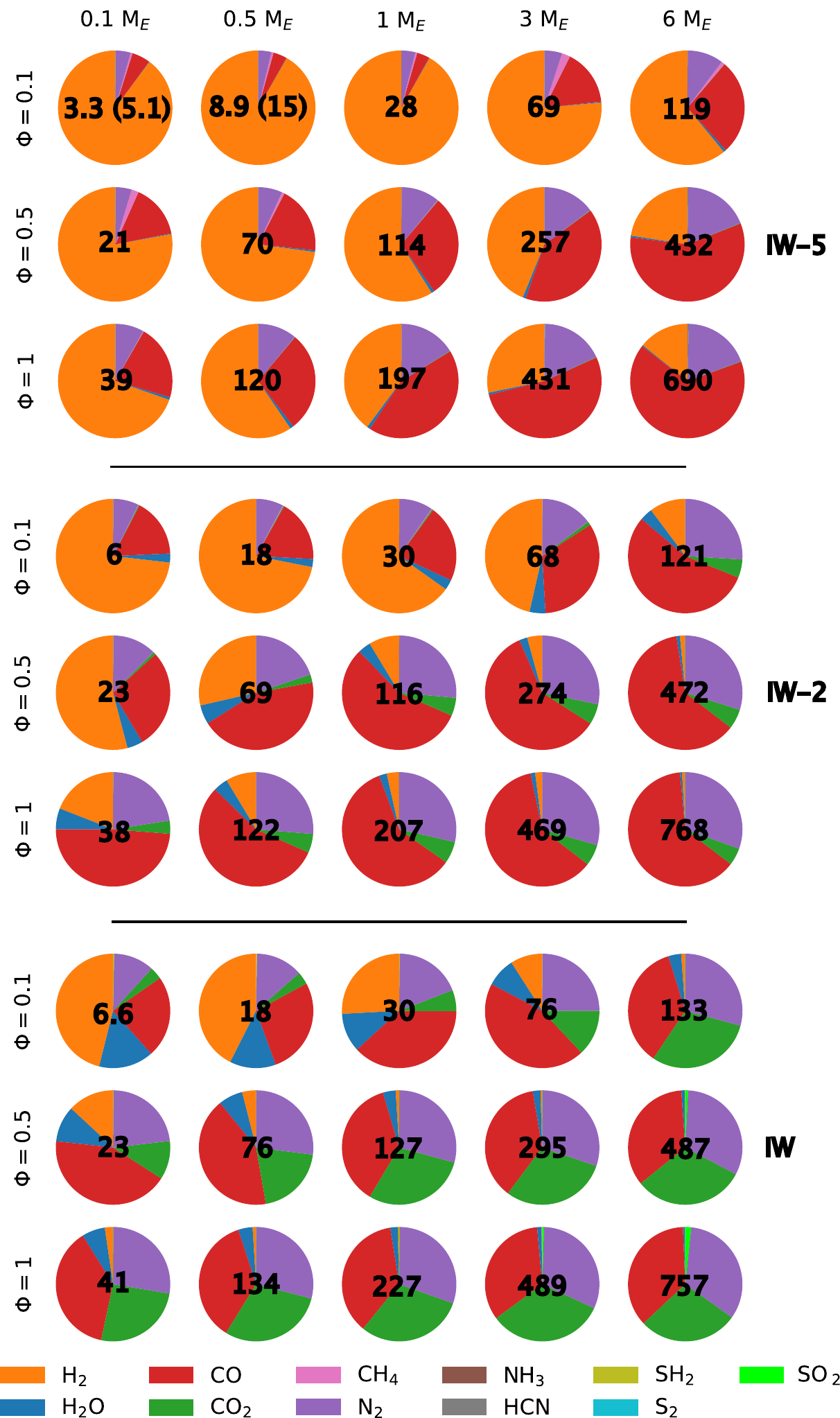}
    \caption{Atmospheric composition (expressed as the molar fraction of each gas in the atmosphere) for exoplanets from 0.1 to 6 Earth masses ($M_E$, columns) and an MO mass from 0.1 to 1 times the planetary silicate mass ($\Phi$, rows) for all three $f_{{\rm O}_2,{\rm eq}}$ investigated (expressed as $\Delta$IW). The numbers indicate the total surface pressure in bar (the number in parentheses indicates the total surface pressure for cases without graphite precipitation). All cases have bulk silicate-equivalent abundances of 100 ppm of H, C, N, and S. The amount of some species (e.g., NH$_3$, HCN, SH$_2$, or S$_2$) is always too small to be noticeable in these charts.}
    \label{fig:pies}
\end{figure}
In this section, we show how variations in the redox conditions provide another important control of the extent of outgassing and the composition of the outgassed atmosphere for a variety of rocky planets in terms of mass. The case of S already illustrated how the decreasing solubility with increasing $f_{{\rm O}_2}$ could dominate the physical controls on outgassing for the deepest MOs and yield an increase in the outgassed mass fraction of S (from $\sim 0.2$ weight percent,  wt\%, of the bulk S budget for Mars-sized planets up to $\>50$ wt\% for 6 Earth masses super Earths; Figure \ref{fig:outgassed_mass_fraction}c and supplementary materials).

Figure \ref{fig:pies} represents the atmospheric composition over the complete parameter space. Two trends are clearly observed: first, the large fraction of H$_2$ at low planetary mass and/or low $\Phi$. This effect, already hinted at in the previous section, is due to the low molecular mass of H$_2$, which amplifies the influence of the planetary mass on its outgassing, in particular, at low $M_p$. Second, at constant $\Delta{\rm IW}_{\rm eq}$, the mixing ratios of the oxidized species systematically increase with increasing planetary mass and $\Phi$. In MO deeper than $6$ GPa (i.e., in this study, all but the shallowest MO on the smallest planet), an increasing depth of the MO causes its surface to become more oxidized \citep{Hirschmann2012,Armstrong2019,Deng2020,Kuwahara2023}, thereby affecting the relative abundances of species in the same redox system. This second effect accentuates the first effect to promote H$_2$ in the atmospheres of small planets and shallow MOs. Over the whole $M_p-\Phi$ space, the surface $\Delta$IW varies by 3 log-units for each value of $\Delta$IW$_{\rm eq}$ (see Figure \ref{fig:profiles}b and supplementary materials), leading to large variations in the atmospheric speciation.

The element H is present in the most gaseous species (6 out of the 11). As a major species in the atmosphere, however, it is essentially carried by H$_2$ and H$_2$O, and it is marginally carried by CH$_4$. The H$_2$-H$_2$O system is dominated by the effect of the low molecular mass of H$_2$ already described and by the high (and non-Henrian) solubility of water. The T-dependence of equilibrium \ref{eq:H_equi} should promote H$_2$O at low temperature (the equilibrium constant increases by $\sim12$ orders of magnitude between the highest and the lowest MO surface temperature investigated here), but it is the opposite of what we observe. This is due to the decrease in surface $f_{{\rm O}_2}$ between these same cases (Figure \ref{fig:profiles}b), by more than 12 orders of magnitude, the decrease in surface $\Delta{\rm IW}$ adds up to the $T-$dependence on the IW buffer, which dominates the variation of the equilibrium constant. From this equilibrium, for an Earth-sized planet with $\Phi\sim0.5$ (corresponding to an MO surface temperature of 2265 K), the fugacities of these two species coincide around IW. At higher oxygen fugacity, water dominates and allows for a very large ingassing of H. At lower oxygen fugacity (see supplementary materials), abundant H$_2$ in the atmosphere allows for the marginal presence (subpercent level) of other hydrogenated species, such as CH$_4$, HCN, NH$_3$, and SH$_2$.

CO is always a major gas in the atmosphere. The importance of CO$_2$ increases with $\Delta$IW$_{\rm eq}$, balancing CO around a surface $f_{{\rm O}_2}$ of IW+1 (for an Earth-sized planet and $\Phi\sim0.5$). CO$_2$ has a higher solubility in silicate melts than CO, thus altering outgassing of C with increasing $f_{{\rm O}_2}$ (i.e., increasing $M_p$, $\Phi$, and/or $\Delta$IW$_{\rm eq}$). Under reducing conditions, the high H$_2$ fugacity allows for some C to be transferred to CH$_4$ and HCN, but always as minor or trace species.  As for the H system, the increase in the equilibrium constant of Equation \ref{eq:C_equi} with decreasing temperature is balanced by that of $f_{{\rm O}_2}$, preventing promotion of CO$_2$ over CO on small planets with a colder MO surface. While graphite saturation in the most reduced cases significantly alters C outgassing, the resulting atmospheric composition is only slightly altered compared to the same cases when this process is not accounted for (although the C/N ratio in the atmosphere is impacted; see Figure \ref{fig:elemental_ratios}). Other species (in particular, H$_2$) are indirectly affected. They maintain comparable molar fractions while the total pressure decreases when C outgassing is suppressed by graphite precipitation.

N is consistently present as N$_2$, but other N carriers such as NH$_3$ and HCN are only marginally present. While  the N solubility strongly depends on $f_{{\rm O}_2}$ and becomes highly soluble under reducing conditions, the N ingassed mass fraction increases only under the most reduced conditions ($\Delta$IW$_{\rm eq}=-5$, $\Phi\leq0.5$ and $M_p\leq1$ Earth mass; Figure \ref{fig:outgassed_mass_fraction}a and b). Over the remaining parameter space, the physical effect of ingassing with increasing planetary mass and molten silicate mass fraction dominates the decreasing solubility of N. This is consistent with the results of \citet{Gaillard2022}, who found N solubility to increase significantly around IW-4. This value is reached (at the surface) only in the most extreme cases of our parameter space (i.e., Mars-size planets, or planets with a shallow MO).  The N redox system is not as T-sensitive as the H and C ones. 

As already pointed out, due to its high solubility at low $f_{{\rm O}_2}$ \citep{Gaillard2022}, S is mostly absent in the atmosphere except in the most oxidized cases, in which it is present as the oxidized end-member SO$_2$, one of the major species of our oxidized super-Earth cases. Here, the increase in the equilibrium \ref{eq:S2_equi} constant with increasing T acts in the same direction as the increase in the $f_{{\rm O}_2}$ on large planets, promoting speciation of S as SO$_2$ and thereby decreasing its effective solubility. In more reduced cases, the presence of H promotes SH$_2$ over S$_2$, which hardly exceeds the ppm level in the atmosphere. S can also precipitate as sulfide under reduced conditions \citep{Blanchard2021}. We tested for it in the same way as we tested for graphite saturation (but in the whole parameter space, sulfide saturation parameterization being calibrated over wider $P$-$T$ conditions), and did not find sulfide precipitation because the S concentration we considered was moderate.

Over the parameter space investigated, lava planets whose oxidation states have been buffered by an event of core formation exhibit a diversity of atmospheric composition, dominated by H for small planets and by C (reduced CO or oxidized CO$_2$) for super-Earths. Nitrogen (as N$_2$) is consistently present, while the presence of of S in the atmosphere is diagnostic of oxidizing conditions. In general, oxidized atmospheres exhibit a larger diversity of major species than reduced atmospheres, which are dominated either by CO and H$_2$ or by H$_2$ alone.

\subsection{Influence of the bulk volatile budget}

The previous results apply for the fiducial case of planets with contents of 100 ppm per mass of MO of each volatile element (H, C, N, and S). We investigated how the observed trends are affected when these budgets were varied. It is extremely difficut to vary the initial budget in volatile elements because it depends on a wide range of processes, from parent body processing in the planetary building blocks to its accretion history. To narrow our parameter space, we focused on two alternative scenarios: 1) an H-enriched case (1000 ppm H), accounting for the possibility of nebular gas capture during the lifetime of the protoplanetary disk (in particular, by massive super-Earths), and 2) an N-depleted case (10 ppm N), as this pattern is a general feature of rocky and icy bodies in the Solar System \citep{Bergin2015}.

Outgassing couplings between the various elements (and their bulk contents) chiefly occur via an alteration of the mean molecular mass of the atmosphere, which modifies the mass distribution between in the MO and the atmosphere. Other couplings, chemical in origin, could in theory involve species bearing more than one element (not counting O), such as CH$_4$, NH$_3$, HCN or SH$_2$. However, the redox and temperature conditions and the high solubility of sulfide strongly disfavor the first three and the last one, respectively.

\begin{figure*}[h!]
    \centering
    \includegraphics[width=\textwidth]{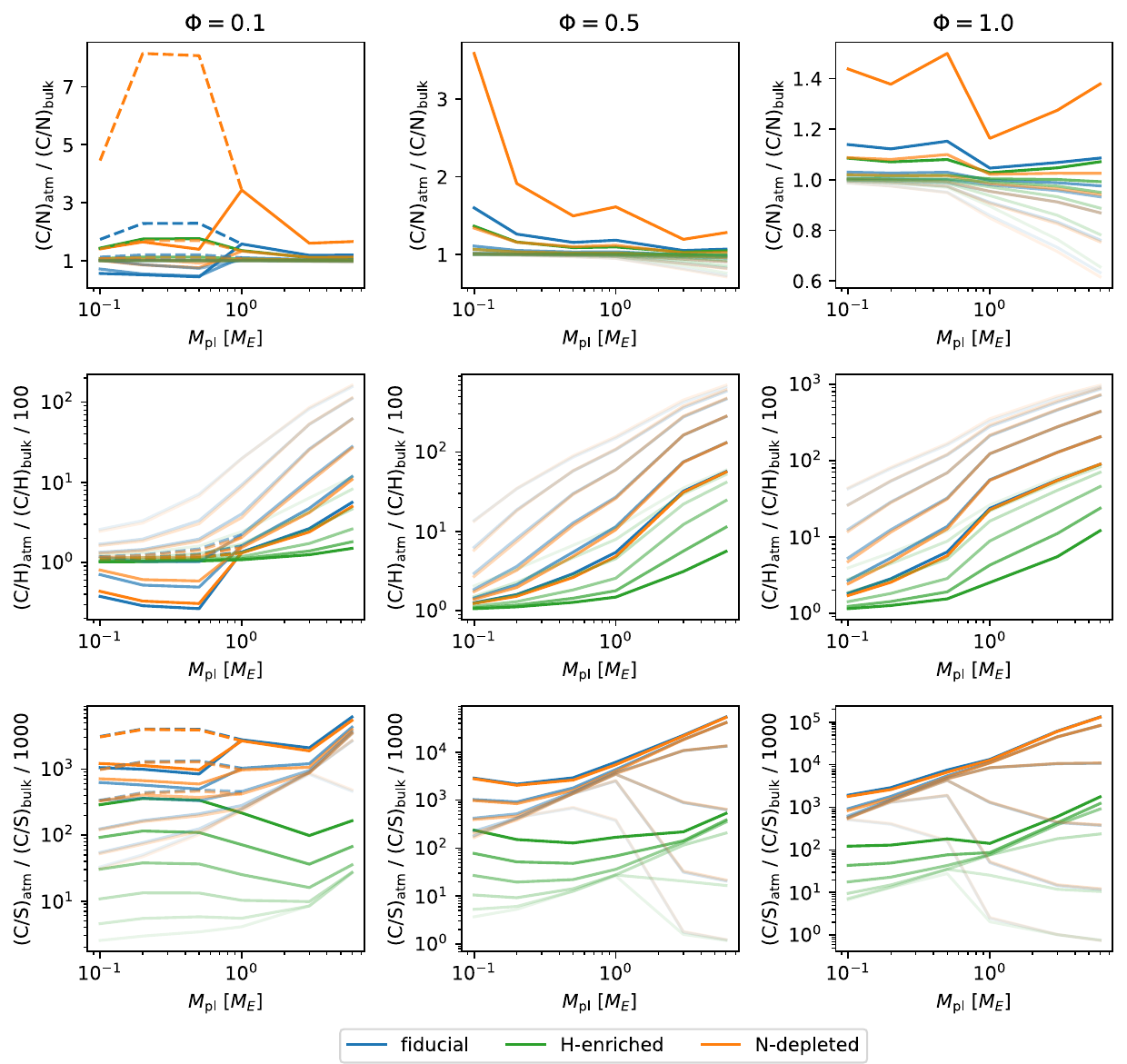}
    \caption{Atmospheric C/N (top row), C/H (middle row), and C/S (bottom row) normalized to the bulk value for all cases with $\Phi=0.1$ (left column), $\Phi=0.5$ (middle column), and $\Phi=1$ (right column). The blue, green, and orange lines correspond to the fiducial, H-rich, and N-depleted cases, respectively. The line shades correspond to the MO redox state (more transparent lines correspond to more oxidized cases, ranging from $\Delta{\rm IW}_{\rm eq}=$-5 to 0). The dashed lines (only in the left column) correspond to cases when graphite saturation was ignored. A bulk-similar atmosphere is plotted along $y=1$, a C-enriched atmosphere (compared to N, H or S) above this line, and an N-, H-, or S-enriched atmosphere (compared to C) is plotted below. The $y$-axis is linear in the top row and logarithmic in the middle and bottom rows.}
    \label{fig:elemental_ratios}
\end{figure*}

Figure \ref{fig:elemental_ratios} represents three elemental ratios (C/N, C/H, and C/S) in the atmosphere, normalized by their bulk value in the MO + atmosphere system. In N-poor cases, N depletion is more pronounced in the outgassed atmosphere than in the bulk MO + atmosphere system (compare the orange and blue lines in the top row of Figure \ref{fig:elemental_ratios}). In the most reduced cases, the enhanced solubility of N  accentuates this effect, which is made obvious when graphite precipitation in the shallowest MOs on the smallest planets is ignored (dashed lines in the top left panel of Figure \ref{fig:elemental_ratios}. The bulk elemental solubility of N increases by a factor 2 between the fiducial case and the N-depleted case, which we attribute to the decrease in mean molecular mass of the atmosphere caused by shortage of N$_2$ (see supplementary materials), and which results in ingassing. These trends affect the N system more strongly due to its decreased abundance, and a divergence from the fiducial case on other elemental ratios is hardly noticeable. The nonmonotonic change in the atmospheric-to-bulk C/N ratio for $\Phi=1$ in the most reduced case (amplified in the N-depleted case) is due to nonlinearities in the chemical system (e.g., the link between surface fO2 and planet size). In the most oxidized cases, C is increasingly speciated as CO$_2$, thereby increasing its effective solubility and decreasing the atmosphere-to-bulk C/N ratio (see the top right panel in Figure \ref{fig:elemental_ratios}).

Interestingly, increasing the H budget suppresses graphite precipitation in the most reduced cases, enforcing a high atmospheric C/N ratio. The N-depleted case is due to the decrease in the mean molecular mass of the atmosphere, this time not due to depletion of heavy species (N$_2$), but to the addition of light ones (H$_2$). Addition of H also significantly reduces atmospheric C/H and C/S, which causes both ratios to become closer to the bulk value (graphite precipitation can decrease the atmospheric C/H below the bulk value in the most reduced cases; see the middle left panel in Figure \ref{fig:elemental_ratios}).
Here again, the addition of H lowers the mean molecular mass in the atmosphere (which always remains below 9 amu in the H-rich cases, but reaches 26 amu in the other cases; see supplementary materials), thereby promoting outgassing of all species and thus causing the atmosphere to become more similar in bulk.

\section{Discussion}
\label{sec:discussion}
We use the results presented above to derive updated mass-radius relations that take the complex chemistry of an outgassed atmosphere into account.
We then discuss the fate of these outgassed atmospheres because potential atmospheric escape can alter the chances of observations as well as the volatile content patterns on exoplanets.

\subsection{Mass-radius relations}
\label{subsec:mass-radius}
Mass-radius relations are crucial theoretical tools for constraining the properties of exoplanets. They are used to infer the internal structure of a planet based on its bulk density. Mass-radius relations for rocky exoplanets have mostly been studied in the light of their refractory composition \citep[e.g.][]{Valencia2006,Dorn2015,Unterborn2016,Unterborn2023,Zeng2016,Zeng2019,Agol2021,Acuna2023}. Exoplanet observations by the transit method measure the transit radius, which, if considered as the actual surface radius of the solid or condensed part of the planet, can lead to a significant error on the planetary structure. Recently, \citet{Turbet2020} showed that the presence of water as steam rather than in a condensed phase could significantly alter the mass-radius relations of terrestrial planets, which are more irradiated than the runaway greenhouse limit, yielding a much larger transit radii. The resulting thick steam atmosphere maintains a high surface temperature, and when the planet is hot enough to melt its silicate envelope, \citet{DornLichtenberg2022} showed that the dissolution of water in the MO could further alter the influence of the steam atmosphere on the mass-radius relation and decrease the transit radius.

\begin{figure*}
    \centering
    \includegraphics[width=\textwidth]{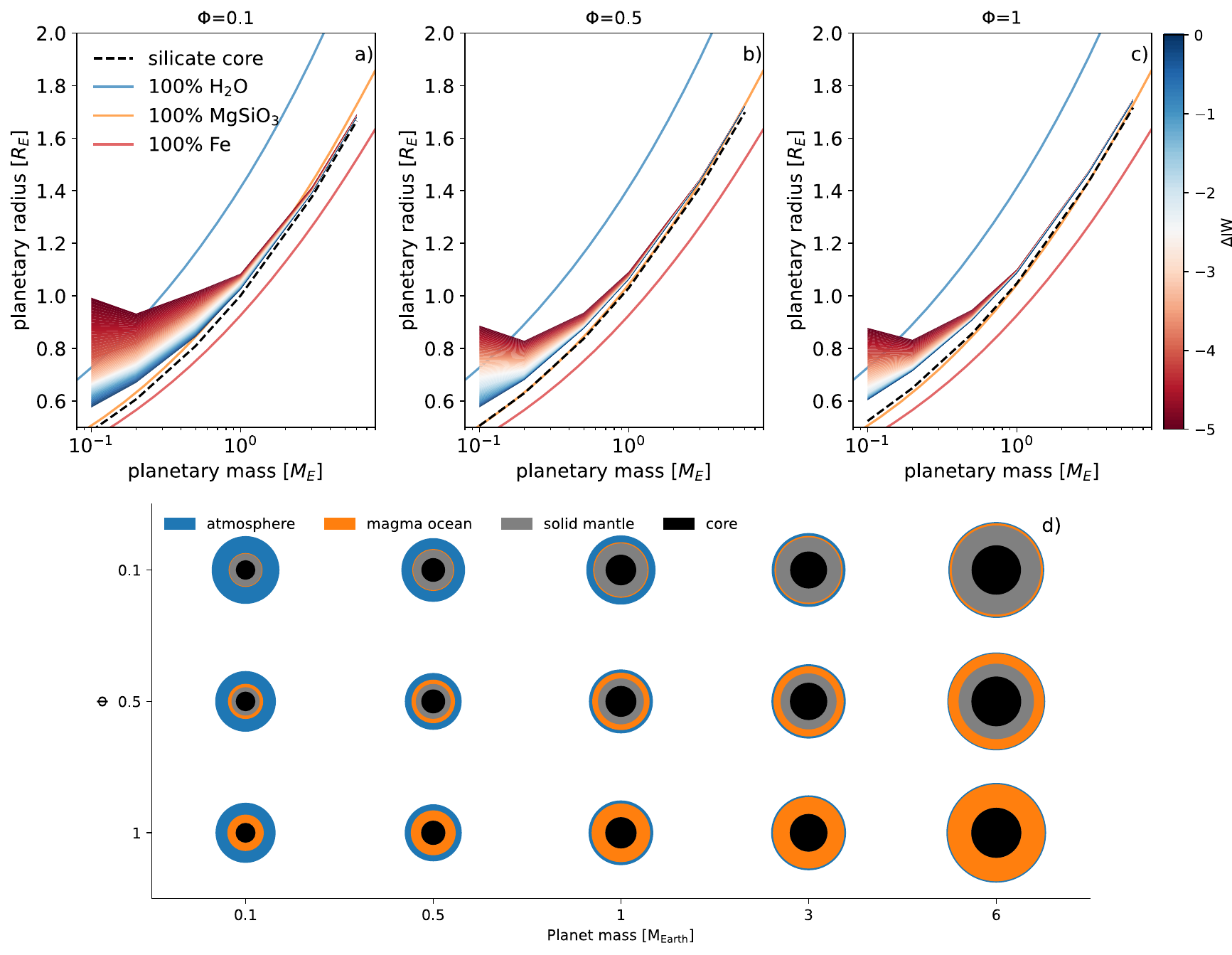}
    \caption{Mass-radius relations of bare silicate core radii (dashed lines) and with the outgassed atmosphere as a function of $\Delta{\rm IW}_{\rm eq}$ (colored contours) for $\Phi=0.1$( a), $\Phi=0.5$ (b), and $\Phi=1$ (c), and relative sizes (to scale) of the solid, liquid, and gaseous part of the planets for the fiducial volatile contents and $\Delta{\rm IW}_{\rm eq}=-5$.}
    \label{fig:mass-radius}
\end{figure*}

While previous studies have focused on steam atmospheres, we showed that outgassing of endogeneous volatiles from an MO can follow very different chemical patterns depending on the planet size, the molten silicate mass fraction, and the redox state (Figure \ref{fig:pies}). In order to assess the influence of these outgassed atmospheres on the mass-radius relations for terrestrial exoplanets, we calculated their transit radii, shown in Figure \ref{fig:mass-radius}a-c (for the fiducial volatile contents). The transit radii were calculated as a function of wavelength using petitRADTRANS \citep{Molliere2019} and were averaged between 0.5 and 1 $\mu$m, which is similar to the wavelength range of most photometric missions, e.g., the CHaracterising ExOPlanets Satellite (CHEOPS), the Transiting Exoplanets Survey Satellite (TESS), the \textit{Kepler} satellite, or the upcoming PLAnetary Transits and Oscillations of stars satellite (PLATO). As observed in Section \ref{subsec:extent_outgassing}, small planets exhibit the strongest relative outgassing, in particular of H$_2$, yielding a large atmospheric vertical extension and thus an observed transit radius that is up to twice as large as the actual planetary radius (Fig. \ref{fig:mass-radius}). For example, a 0.1 Earth-mass planet could appear to be as large as an Earth-sized planet (\ref{fig:mass-radius}a and d). Interestingly, graphite precipitation on a planet like this, although decreasing the surface pressure, also decreases the mean molar mass of the atmosphere, yielding larger vertical extension (see Table \ref{tab:planetary_cores}).

Changing the volatile budgets by either decreasing N or increasing H also affects the mean molecular mass of the atmosphere. While the former only has a negligible effect, the latter (increasing the H budget by one order of magnitude) leads to significantly inflated super-Earths, whose bulk densities are similar to a 100\% H$_2$O composition. No realistic transit radii for H-rich reduced small planets could be obtained because the assumption of hydrostatic equilibrium breaks down in these conditions of surface pressure, temperature, and gravity. Furthermore, we note that our mass-radius relations for the refractory part of the planet yield similar densities for whole-mantle MO planets ($\Phi=1$) and 100 \% (solid) MgSiO$_3$ planets from \citet{Zeng2016} (Figure \ref{fig:mass-radius}c), which is partly due to the presence of an MO (less dense than cold mantle), and partly due to potential differences between the equations of state from burnman and those used in \citet{Zeng2016}. While this is likely an underestimate of the bulk density, it is still dominated by the effect of the atmospheres for small planets and for H-enriched super-Earths.

The very low apparent density of small reduced exoplanets (an overestimate by 100\% of the radius corresponding to an eightfold underestimate of the density) is a striking feature that might be mistaken for a nebular accreted primordial atmosphere, however. Nevertheless, as an MO-causing giant impact can occur much later than disk dissipation \citep{Raymond2004}, and considering the likely limited survival time of H$_2$-rich atmospheres around small planets, an outgassed origin would provide an explanation in case a small, low-density planet were observed around a relatively young star (a few 100 Myr old).
The transit depth (estimated as $\left(R_{\rm tr}/R_{\rm st}\right)^2$, where $R_{\rm st}$ is the stellar radius) of the most reduced 0.1 $M_E$ planets is $~80$ ppm around a Sun-like star. This is similar to the detection limit of the upcoming PLATO mission\cite{PLATO}. \citet{Lupu2014} and \citet{Bonati2019} made the case for the search of transient lava worlds in young systems, which our result helps to understand.

\subsection{Fate of secondary atmospheres}
\label{subsec:fate_volatiles}
\begin{figure}[h!]
    \centering
    \includegraphics[width=9cm]{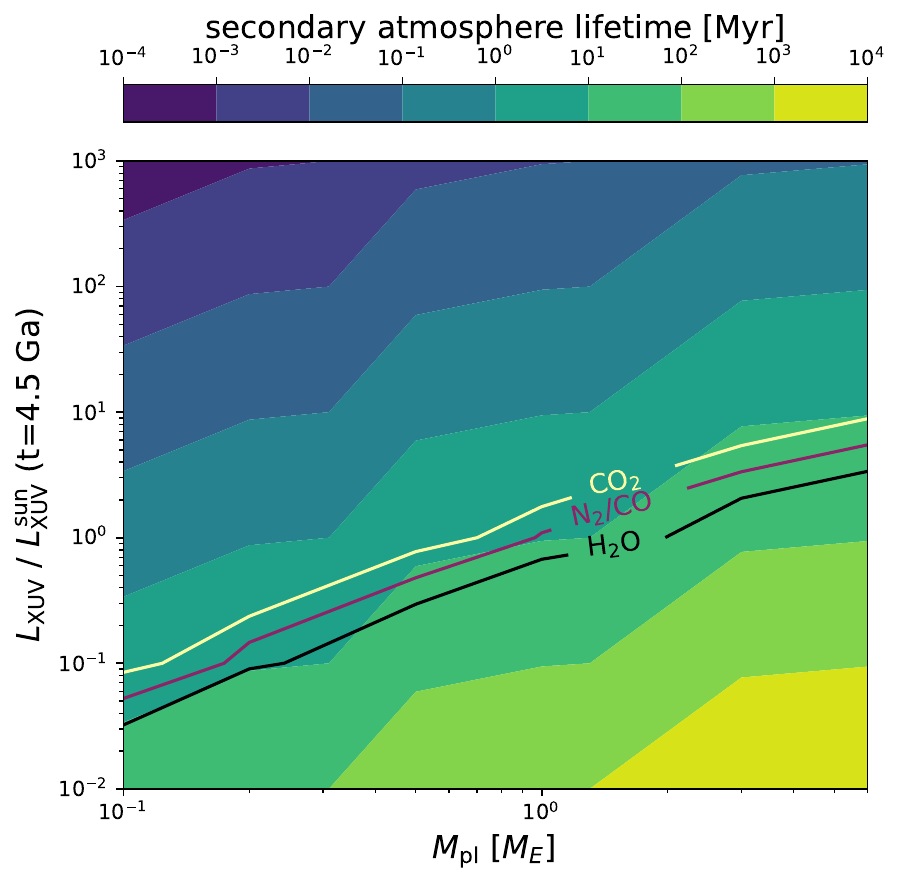}
    \caption{Lifetime (in Myr) of the MO-outgassed atmosphere as a function of planetary mass ($x$-axis), and incoming XUV flux ($L_{\rm XUV}$, $y$-axis), normalized by the solar XUV flux received by the Earth 4.5 Ga ($L_{\rm XUV}^{\rm sun}$(t=4.5 Ga)) from \citet{Ribas2005}, for the cases most prone to hydrodynamic escape ($\Delta{\rm IW}=-5$ and $\Phi=0.1$). The loci in which the crossover mass matches the molecular mass of H$_2$O (black), CO or N$_2$ (purple), and CO$_2$ (white) are also represented.}
    \label{fig:atm_lifetime}
\end{figure}

After they are outgassed, volatiles are prone to atmospheric escape. H$_2$-rich inflated atmospheres are particularly sensitive to hydrodynamic escape, where heating of the upper atmosphere results in a pressure gradient that destabilizes hydrostatic equilibrium and induces an outward flow. The gas is advected outside of the gravitational zone of influence of the planet (the Bondi or Hill sphere), and is lost. An upper limit for the mass-loss rate is provided by the energy-limited regime, where the entirety of the stellar extreme-UV (XUV) flux received by the planet is used to lift the atmosphere \citep{Watson1981},
\begin{equation}
    F_{{\rm H}_2}=\frac{\pi R_{\rm XUV}^2f_{\rm XUV}\epsilon R_p}{GM_p},
    \label{eq:hydrodynamic_escape}
\end{equation}
where $F_{{\rm H}_2}$ is the H$_2$ escape flow (in kg/s), $R_{\rm XUV}$ is the XUV absorption radius, which we took to be equal to the transit radius calculated in the previous section, $f_{\rm XUV}$ is the incoming XUV flux (in W/m$^2$), $\epsilon$ is an efficiency coefficient, which we took to be equal to 0.3 following \citet{Katyal2020}, and $G$ is the universal gravitational constant. Assuming energy-limited H$_2$ escape, Figure \ref{fig:atm_lifetime} represents the lifetime of an outgassed atmosphere as a function the planetary mass and the incoming XUV flux for the most reduced cases ($\Delta{\rm IW}=-5$) and $\Phi=0.1$. While Mars-sized planets lose their H$_2$-dominated atmospheres in less than 1 Myr under most instellation conditions for young stars, H$_2$-dominated secondary atmospheres on larger planets that also exhibit anomalously large transit radii can be retained over several million years for early-Earth-like XUV instellations. Furthermore, the presence of heavier species (already at the percent level) reduces the efficiency of hydrodynamic H loss by several orders of magnitude on Mars- \citep{YoshidaKuramoto2020} and Earth-sized planets \citep{YoshidaKuramoto2021}, provided that these species are infrared emitters. In particular, \citet{YoshidaKuramoto2020} showed that an H$_2$ atmosphere of 6 bar on Mars with 7\% CO-CH$_4$ (comparable to our most reduced case) can be retained for $\sim8$ Myr due to the radiative cooling of C-bearing species that compete with XUV heating and decrease the escape efficiency. 

Escape of H$_2$-rich atmospheres can carry diluted heavier species and trigger loss of elements other than H \citep{KiteBarnett2020}. This occurs for species lighter than the crossover mass \citep{Hunten1987}, which is defined as
\begin{equation}
    \mu_{\rm crossover}=\mu_{{\rm H}_2}+\frac{kTF_{{\rm H}_2}}{bgX_{{\rm H}_2}},
    \label{eq:crossover_mass}
\end{equation}
with $k$ the Boltzmann constant, $T$ the temperature in the upper atmosphere (set to the stratospheric temperature, 200 K), $b=2.2\times10^{21}$ m$^{-1}$.s$^{-1}$ the diffusion parameter \citep{Hunten1987}, and $X_{{\rm H}_2}$ the molar mixing ratio of H$_2$ in the atmosphere. In Figure \ref{fig:atm_lifetime}, we also indicate by contours where the crossover mass matches the molecular mass of H$_2$O, CO, or N$_2$, and CO$_2$. Above these lines, the respective species escapes along with H$_2$, and hydrodynamic escape affects the bulk volatile content of their constitutive elements. However, we stress again that most of these species (except for N$_2$) are strong infrared emitters, can cool the upper atmosphere, and  can significantly decrease the escape rate. Furthermore, it is questionable whether the most massive super-Earths, whose atmosphere is not dominated by H$_2$ (Figure \ref{fig:pies}, third last row), are in the hydrodynamic escape regime. Only an accurate modeling of hydrodynamic escape of MO atmospheres, which is beyond the scope of this study, can be conclusive on the escape of the various volatile elements during this phase.

Massive loss of H$_2$ can also alter the redox state of the planet because H loss corresponds to a net oxidation \citep{Sharp2013,Wordsworth2018,KiteSchaefer2021}. In this case, a new $f_{{\rm O}_2}$ forcing sets in, offsetting the FeO$-$Fe$_2$O$_3$ buffer toward a higher $f_{{\rm O}_2}$. \citet{Sharp2013} calculated that the loss of 1.3 Earth water-ocean-equivalent of H (corresponding to 455 ppm H$_2$O of the BSE mass, or 50 ppm H) would raise the $f_{{\rm O}_2}$ of Earth's mantle from IW to FMQ, which is an increase of $\sim4$ log-units. Assuming that we can extrapolate the trend to lower $f_{{\rm O}_2}$ values, this corresponds to a crossing of the complete $f_{{\rm O}_2}$ space we investigated upon losing half of the H endowment. This could raise the initially most reduced cases to the most oxidized ones as a result of H escape. With increasing $f_{{\rm O}_2}$, H$_2$ is converted into H$_2$O, which is heavier and dissolves more readily in the magma, thus shutting down atmosphere escape. It is therefore unlikely that even the smallest reduced exoplanets become fully dessicated.

\section{Conclusions}
\label{sec:conclusions}
By investigating the atmospheric composition and extent of MO outgassing on exoplanets from 0.1 to 6 Earth masses, with molten silicate mass fractions from 0.1 (shallow MO) to 1 (whole-mantle MO), and with core-mantle equilibration $f_{{\rm O}_2}$ between IW-5 and IW (yielding to MO-atmosphere eaquilibrium between IW-5.5 and IW+1.5), we reached the conclusions that we list below.
\begin{enumerate}
    \item The extent of outgassing decreases with increasing planetary mass and MO depth. The low molecular mass of H$_2$ amplifies this mechanism, leading to strong outgassing of H (as H$_2$) on small planets and shallow MOs. 
    \item The MO and atmosphere oxidation increases with increasing planetary mass and MO depth,
increasing H and C retention in MO, but decreasing N and S retention.
    \item C and N are mostly in the atmosphere, while H and S are mostly dissolved, except under the most oxidized conditions, where SO$_2$ is outgassed, and the most reduced conditions, where atmospheres become consistently H$_2$ rich. Graphite saturation in the most reduced cases can also drastically decrease C outgassing.
    \item Extensive H outgassing on reduced small planets leads to a high vertical extension of their atmospheres, significantly affecting their mass-radius relations. This effect is amplified by limited C outgassing due to graphite saturation.
    \item Extensive H outgassing on small planets makes them prone to H$_2$ escape, while volatiles are efficiently retained in the oxidized atmospheres with high molecular mass of super Earths (in addition to their increased gravity).
\end{enumerate}
These results should inform future atmosphere-retrieval models applied to potential young lava worlds. In particular, several of the species predicted by our model, exhibiting strong absorption features in the infrared, offer an interesting target for spectroscopic studies.

\begin{acknowledgements}
    The authors thank the anonymous reviewer for their constructive comments that helped improve the manuscript. M. M. thanks Tatsuya Yoshida for insightful discussions on hydrodynamic escape.
    This work received funding from the NASA grant 80NSSC18K0828.
\end{acknowledgements}

%
%

\appendix

\section{Solubility laws}

\begin{table*}[]
    \begin{minipage}{\textwidth}
    \centering
    \caption{Species-specific gas solubility laws (above the line) and graphite/diamond and sulfide saturation parameterizations (below the line).}
    \begin{tabular}{l c c c c c}
        \bf{Species} & \bf{Reference} & \bf{Temperature [°C]} & \bf{Pressure [bar]} & \bf{O$_2$ fugacity [-]} & \bf{Composition}\\ \hline
        H$_2$  & \cite{Hirschmann2012} & 1400-1450 & 7000-30000 & IW & basalt, andesite \\
        H$_2$O & \cite{Sossi2023} & 1900 & 1 & IW$-$1.9-IW$+$6 & peridotite \\
        CH$_4$    & \cite{Hirschmann2016} & 1400-1800 & 350-32000 & IW$-$3.6-IW$+$2.3 & basalt \\
        CO     & \cite{Hirschmann2016} & 1400-1800 & 350-32000 & IW$-$3.6-IW$+$2.3 & basalt \\
        CO$_2$    & \cite{Hirschmann2016} & 1200-1625 & 100-25000 & IW$-$1-FMQ$+$2 & basalt \\
        N$_2$     & \cite{Dasgupta2022} & 1050-2327 & 1-82000 & IW$-$8.3-IW$+$8.7 & basalt \\
        NH$_3$\footnote{Dissolution of these species has been neglected.\label{refnote1}}    & -- & -- & -- & -- & -- \\
        HCN\ref{refnote1}    & -- & -- & -- & -- & -- \\
        SH$_2$\ref{refnote1}    & -- & -- & -- & -- & -- \\
        S$_2$     & \cite{Gaillard2022} & 800-1350 & 1-2000 & IW$-$8.3-IW$+$5 & basalt, dacite\\
        SO$_2^a$    & -- & -- & -- & -- & -- \\ \hline
        CO     & \cite{Yoshioka2019} & 1200-1600 & 2000-30000 & NA\footnote{Not Available.} & MORB, andesite, rhyolite \\
        CO$_2$ & \cite{EguchiDasgupta2018} & 950-1600 & 500-30000 & FMQ$-$6-FMQ$+$1.5 & foidite, rhyolite \\
    \end{tabular}
    \label{tab:solubilities}
    \end{minipage}
\end{table*}

Table \ref{tab:solubilities} lists the solubilitiy parameterizations we used, as well as their temperature, pressure, and $f_{{\rm O}_2}$ calibration ranges, when available. In addition to the experiments presented in these studies, the parameterizations we derived rely on a collection of data from the literature. The ranges indicated in Table \ref{tab:solubilities} reflect this as accurately as possible. We refer to the studies cited in these references for more information.

\section{Mass-radius relations}

\begin{table*}
    \begin{minipage}{\textwidth}
    \centering
    \caption{Radii of the various refractory reservoirs and transit radii and atmospheric pressures for the synthetic lava planet set.}
    \begin{tabular}{c c|c c c c |c c|c c|c c}
        & & & & & & IW-5 & & IW-2 & & IW & \\\cline{7-12}        
        $M_p$\footnote{Masses are expressed in Earth masses ($M_E$), radii are expressed in Earth radii ($R_E$), temperatures are expressed in K, and pressures in bar.\label{refnote3}} & $\Phi$ & $R_{\rm{CMB}}$\ref{refnote3} & $R_{\rm{MO}}$\ref{refnote3} & $R_{\rm{pl}}$\ref{refnote3} & $T_{\rm sfc}$\ref{refnote3} & $R_{\rm tr}$\ref{refnote3} & $P_{\rm atm}$\ref{refnote3} & $R_{\rm tr}$\ref{refnote3} & $P_{\rm atm}$\ref{refnote3} & $R_{\rm tr}$\ref{refnote3} & $P_{\rm atm}$\ref{refnote3} \\ \hline
        0.1 & 0.1 & 0.25 & 0.47 & 0.50 & 1415 & 0.99 (0.72)\footnote{Values obtained when graphite saturation was not taken into consideration.\label{refnote2}} & 3.3 (5.1)\ref{refnote2} & 0.65 & 6 & 0.57 & 6.6 \\
            & 0.5 & 0.25 & 0.40 & 0.52 & 1613 & 0.88 & 21 & 0.65 & 23 & 0.57 & 23 \\
            & 1 & 0.25 & 0.25 & 0.54 & 1885 & 0.87 & 39 & 0.63 & 38 & 0.60 & 41 \\ \hline
        0.5 & 0.1 & 0.40 & 0.79 & 0.83 & 1528 & 1.01 (0.92)\ref{refnote2} & 8.9 (15)\ref{refnote2} & 0.88 & 18 & 0.84 & 18 \\
            & 0.5 & 0.41 & 0.67 & 0.85 & 2019 & 0.93 & 70 & 0.88 & 69 & 0.87 & 76\\
            & 1 & 0.41 & 0.40 & 0.87 & 2440 & 0.94 & 120 & 0.91 & 122 & 0.90 & 134 \\ \hline
        1 & 0.1 & 0.49 & 0.97 & 1.02 & 1636 & 1.08 & 28 & 1.04 & 30  & 1.02 & 30 \\
          & 0.5 & 0.49 & 0.82 & 1.05 & 2265 & 1.09 & 114 & 1.06 & 116 & 1.06 & 127 \\
          & 1 & 0.50 & 0.50 & 1.07 & 2716 & 1.11 & 197 & 1.09 & 207 & 1.08 & 227 \\ \hline
        3 & 0.1 & 0.64 & 1.33 & 1.40 & 1942 & 1.41 & 69 & 1.39 & 68 & 1.38 & 121 \\
          & 0.5 & 0.65 & 1.10 & 1.43 & 2703 & 1.44 & 257 & 1.43 & 274 & 1.43 & 472 \\
          & 1 & 0.65 & 0.65 &  1.45 & 3274 & 1.47 & 431 & 1.46 & 469 & 1.46 & 768  \\ \hline
        6 & 0.1 & 0.75 & 1.60 & 1.69 & 2221 & 1.68 & 119 & 1.67 & 121 & 1.67 & 133 \\
          & 0.5 & 0.75 & 1.30 & 1.71 & 2960 & 1.72 & 432 & 1.72 & 472 & 1.71 & 487 \\
          & 1 & 0.77 & 0.81 & 1.76 & 4229 & 1.74 & 690 & 1.74 & 768 & 1.74 & 757 \\
    \end{tabular}
    \label{tab:planetary_cores}
    \end{minipage}
\end{table*}

Table \ref{tab:planetary_cores} lists the mass-radius relations for the silicate cores of planets that were calculated as described in Section \ref{subsec:mass-radius}, and the transit radii and atmospheric pressures.

\section{Derivation of the mass-conservation equation}
\label{apx:mass_cons}
Let $e$ be an element of atomic mass $\mu_e$ whose mass is conserved, and $s$ be a gaseous species of molecular mass $\mu_s$, composed of (among others) $\lambda_s^e$ atoms of $e$. If $p_s$ is the partial pressure of $s$ in the atmosphere and $P$ and $\mu_{\rm atm}$ are the total atmospheric pressure and mean molecular mass, respectively, then the molar fraction of $s$ in the atmosphere is $p_s/P$ and its mass fraction is $p_s/P\times \mu_s/\mu_{\rm atm}$. Since the total mass of the atmosphere (under the assumption of hydrostatic equilibrium) is its the mass of $s$ in the atmosphere is $4\pi R_p^2/g$, then the mass of $s$ in the atmosphere is
\begin{equation}
    M_s^{\rm atm}=\frac{4\pi R_p^2}{g}p_s\frac{\mu_s}{\mu_{\rm atm}}.
    \label{eq:M_atm_s}
\end{equation}

Let $S_s$ be the solubility law of species $s$ (in general, a function of $p_s$, $P$, and $f_{{\rm O}_2}$. Assuming that the atmosphere and the MO are in equilibrium (see \citet{SalvadorSamuel2023} for a discussion of the validity of this assumption)), the mass of $s$ dissolved in the MO is simply
\begin{equation}
    M_s^{\rm MO}=M_{\rm MO}S_s(p_s,P,f_{{\rm O}_2}).
    \label{eq:M_MO_s}
\end{equation}

The total mass of $s$ in the system is the sum of \ref{eq:M_MO_s} and \ref{eq:M_atm_s}, and the mass of $e$ carried by molecules of $s$ is the mass of $s$ multiplied by $(\mu_e/\mu_s)\lambda_s^e$. Summing over all species (whose masses are not conserved because of chemical reactions), we obtain Equation \ref{eq:mass_cons}).

\section{Benchmark of the chemical equilibrium model}
\label{apx:benchmark}
The chemical equilibrium model we used relies on a number of parameterizations for gas solubilities, gas-gas equilibria, and oxygen fugacity buffers. These choices, in addition to numerical implementation, contribute to potential discrepancies with other models. In order to assess these discrepancies, we reproduced the cases presented in Figures 4 and 8 of \citet{Bower2022} using our model as well as the model of \citet{Gaillard2022} (see supplementary materials). These cases (corresponding to a whole-mantle MO on an Earth-sized planet) cover a wide range of $f_{{\rm O}_2}$ (from IW$-2$ to IW$+4$ at the surface) and of H and C bulk contents (from 39 to 389 ppm, and from 0 to 1944 ppm, respectively). \citet{Bower2022} considered a C-O-H system, and we therefore used a negligible bulk N and S content ($10^{-4}$ ppm). We only reproduced the initial state, as the final state (also presented in these Figures) has a volatile element content that is affected by MO crystallization, which is not reported.

We find that the results of all three models agree in terms of orders of magnitude and C/H outgassing. We note a systematic offset in the reduced-to-oxidized species ratio (e.g., H$_2$/H$_2$O for the H system) between our model and that of \citet{Gaillard2022} on one hand and the model \citet{Bower2022} on the other hand, the latter having higher ratios at the same $f_{{\rm O}_2}$. This discrepancy is likely due to the difference in the IW buffer used by \citet{Bower2022}, from \citet{ONeillEggins2022}, which differs by 0.5 log units from the model we used. The model of \citet{Gaillard2022} only converged for $\Delta{\rm IW}<0$. Importantly, while this exercise illustrates agreements and discrepancies between models, the results are still only as good as the extrapolations of the parameterizations in use outside of the calibration ranges (see Table \ref{tab:solubilities}).

\end{document}